\documentclass[letterpaper]{article} 
\usepackage{aaai2026}  
\usepackage{times}  
\usepackage{helvet}  
\usepackage{courier}  
\usepackage[hyphens]{url}  
\usepackage{graphicx} 
\usepackage{natbib}  
\usepackage{caption} 
\usepackage{algorithm}
\usepackage{algorithmic}

\usepackage{pifont}
\usepackage{amsmath}
\usepackage{amsfonts} 
\usepackage{booktabs}
\usepackage{multirow}
\usepackage{subcaption}

\usepackage{xcolor} 

\usepackage{newfloat}
\usepackage{listings}
\DeclareCaptionStyle{ruled}{labelfont=normalfont,labelsep=colon,strut=off} 
\floatstyle{ruled}
\newfloat{listing}{tb}{lst}{}
\floatname{listing}{Listing}
\title{Diversity Recommendation via Causal Deconfounding of Co-purchase Relations and Counterfactual Exposure}
\author {
    Jingmao Zhang\textsuperscript{\rm 1}\equalcontrib,
    Zhiting Zhao\textsuperscript{\rm 1}\equalcontrib,
    Yunqi Lin\textsuperscript{\rm 1}\equalcontrib,
    Jianghong Ma\textsuperscript{\rm 1}\thanks{Corresponding author.},
    Tianjun Wei\textsuperscript{\rm 2}\thanks{Corresponding author.},
    Haijun Zhang\textsuperscript{\rm 1},
    Xiaofeng Zhang\textsuperscript{\rm 1}
}
\affiliations {
    \textsuperscript{\rm 1}Harbin Institute of Technology(Shenzhen), Shenzhen, China\\
    \textsuperscript{\rm 2}Nanyang Technological University, Singapore\\
    220810426@stu.hit.edu.cn, 24s151185@stu.hit.edu.cn, 210110519@stu.hit.edu.cn,\\
    majianghong@hit.edu.cn, tjwei2-c@my.cityu.edu.hk, hjzhang@hit.edu.cn, zhangxiaofeng@hit.edu.cn
}

\begin{document}

\maketitle

\section{EXPERIMENT SUPPLEMENT}
\subsection{Source Code}The source code has been made anonymously available via https://anonymous.4open.science/status/AAAI-BF40. 
\subsection{Implementation details}
We implemented Cadence using PyTorch and trained it with the Adam optimizer, setting the learning rate to 0.001. The batch size was fixed at 2048, and the embedding dimension was set to 32. For the UACR-driven item-item aggregation layer, we set $L_{II} = 2$. The number of propagation layers in LightGCN was set to $L = 3$. For the two-stage candidate selection in the CSCE module, the global selection number $K_g$ was set to 4 for Beauty, 12 for TaoBao, and 6 for Toy, while the category-specific selection number $K_c$ was set to 1 across all three datasets. The scaling factor $\alpha$ was set to 1.15 for Beauty, 1.05 for TaoBao, and 1.15 for Toy. 


We employ the Bayesian Personalized Ranking (BPR) \cite{rendle2012bpr} loss with negative sampling for training. The BPR loss is defined as:
\begin{equation}
\mathcal{L}_{BPR} = \sum_{(u,i,j) \in D_S} -\ln \sigma(r_{u,i} - r_{u,j}) + \lambda ||\boldsymbol{\theta}||^2
\end{equation}
where $D_S$ is the set of training triplets, $\sigma$ is the sigmoid function, $r_{u,i}$ and $r_{u,j}$ are predicted scores for positive item $i$ and negative item $j$, $\lambda$ is the regularization coefficient, and $\boldsymbol{\theta}$ represents the model parameters. 

\subsection{Computing Infrastructure}
All experiments were conducted using NVIDIA A100 80GB PCIe GPUs on an Ubuntu 22.04 system, with Python 3.9 and PyTorch version 2.4.

\subsection{Parameter analysis}

\begin{figure}[htbp]
\centering
\begin{subfigure}[b]{0.5\columnwidth}
    \centering
    \includegraphics[width=\textwidth]{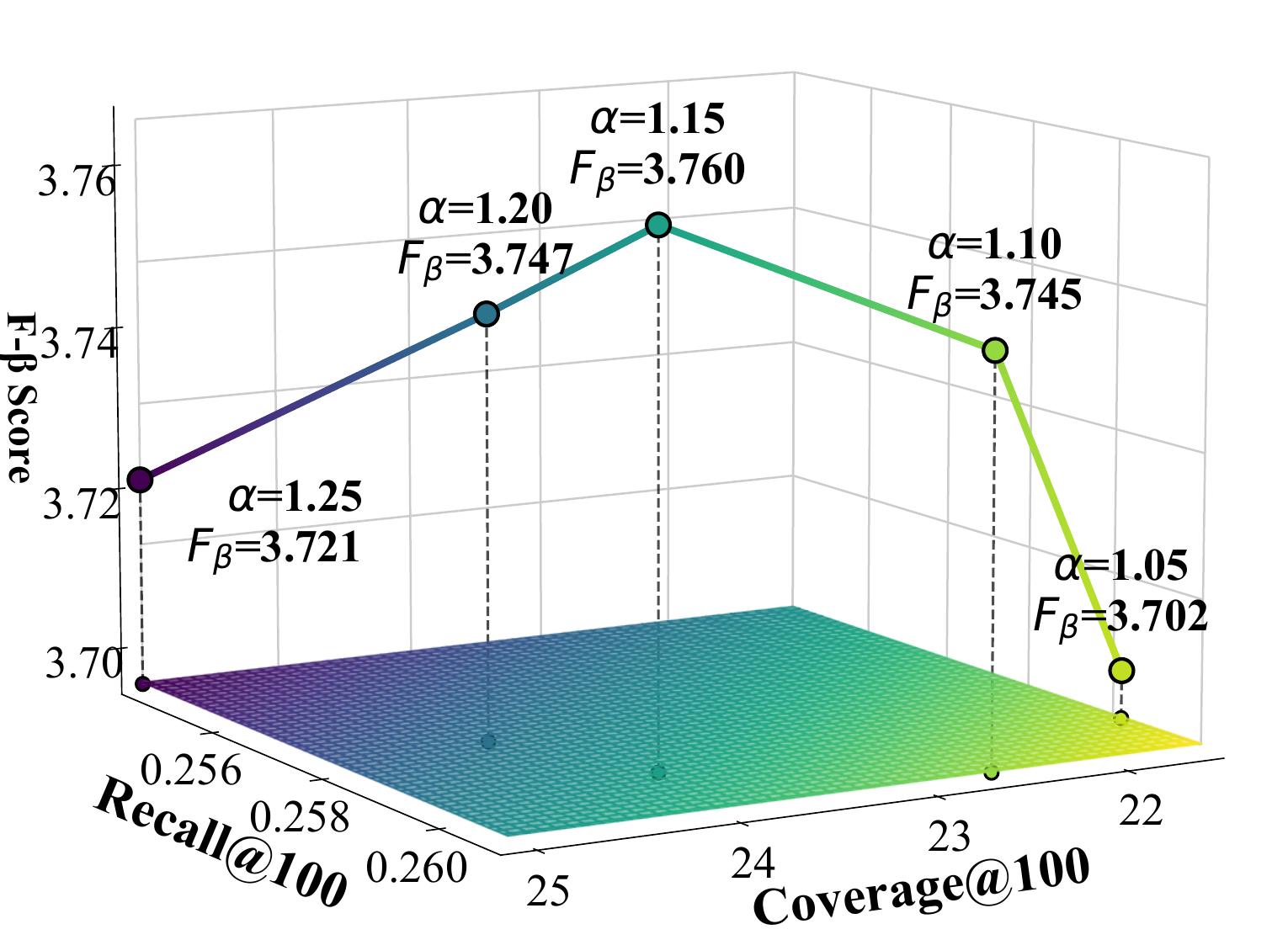}
    \caption{Beauty-$\alpha$}
    \label{fig:beauty_alpha}
\end{subfigure}%
\hfill%
\begin{subfigure}[b]{0.5\columnwidth}
    \centering  
    \includegraphics[width=\textwidth]{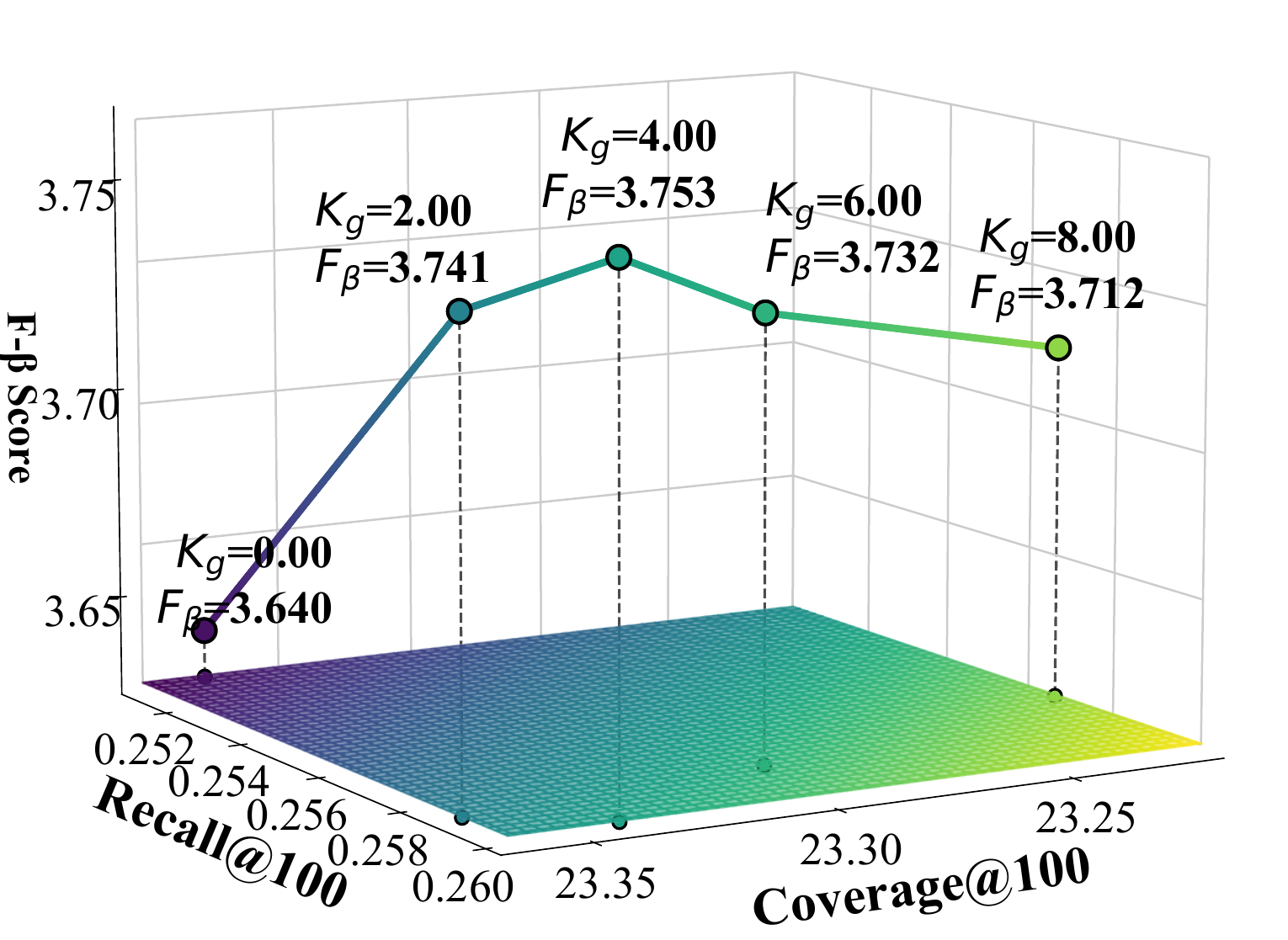}
    \caption{Beauty-$K_g$}
    \label{fig:beauty_kg}
\end{subfigure}

\vspace{0.3em}

\begin{subfigure}[b]{0.5\columnwidth}
    \centering
    \includegraphics[width=\textwidth]{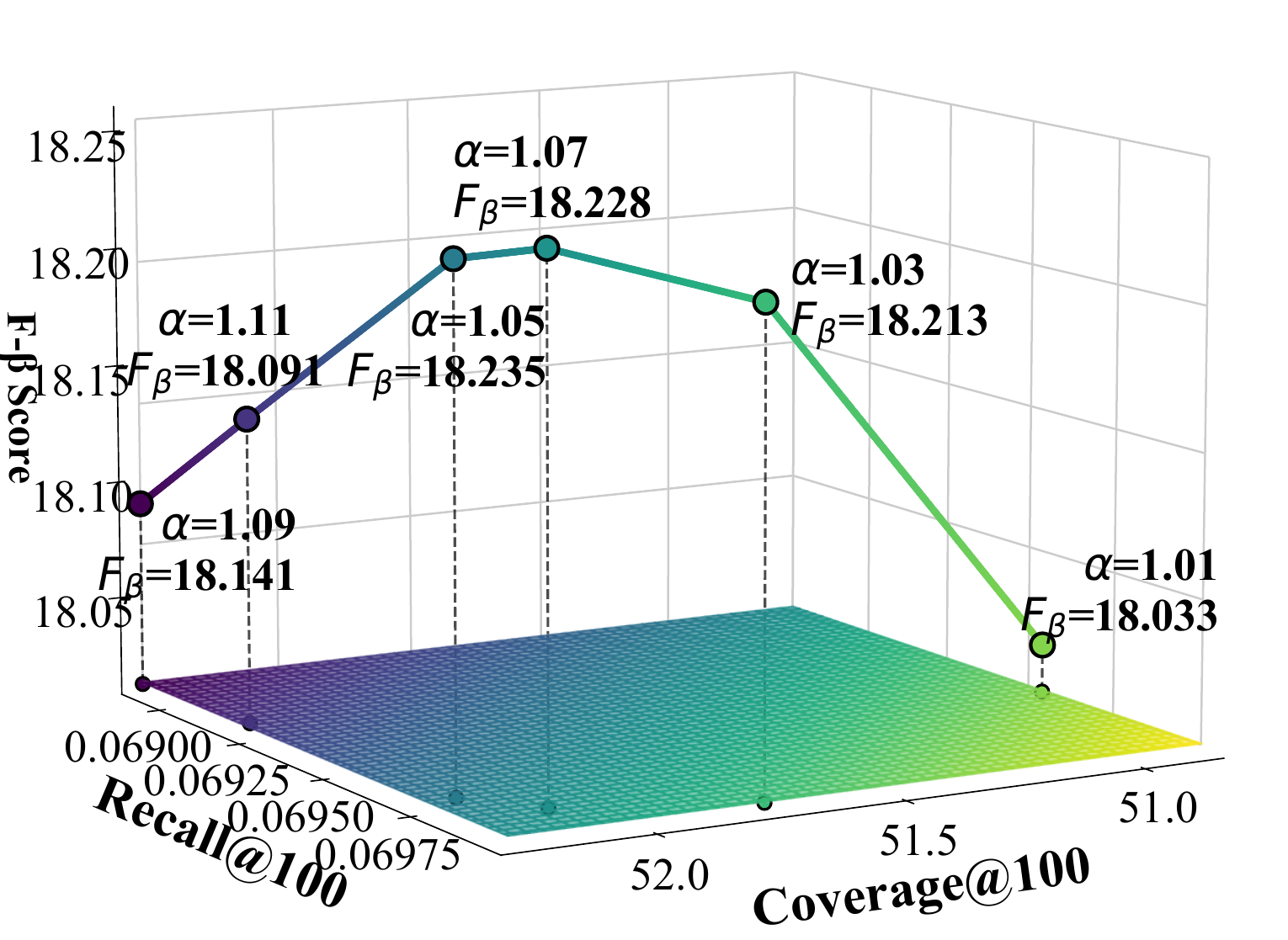}
    \caption{TaoBao-$\alpha$}
    \label{fig:taobao_alpha}
\end{subfigure}%
\hfill%
\begin{subfigure}[b]{0.5\columnwidth}
    \centering
    \includegraphics[width=\textwidth]{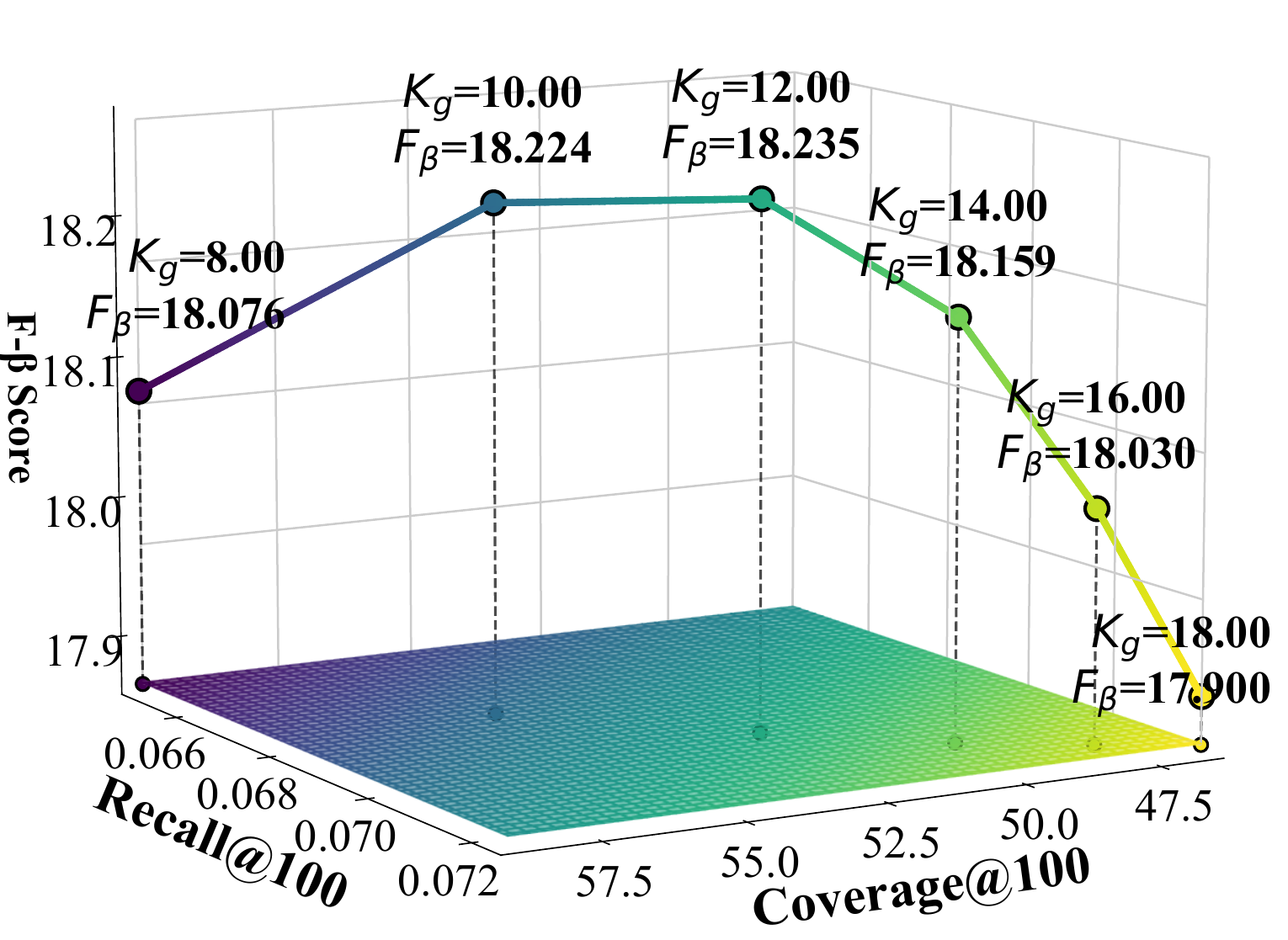}
    \caption{TaoBao-$K_g$}
    \label{fig:taobao_kg}
\end{subfigure}

\vspace{0.3em}

\begin{subfigure}[b]{0.5\columnwidth}
    \centering
    \includegraphics[width=\textwidth]{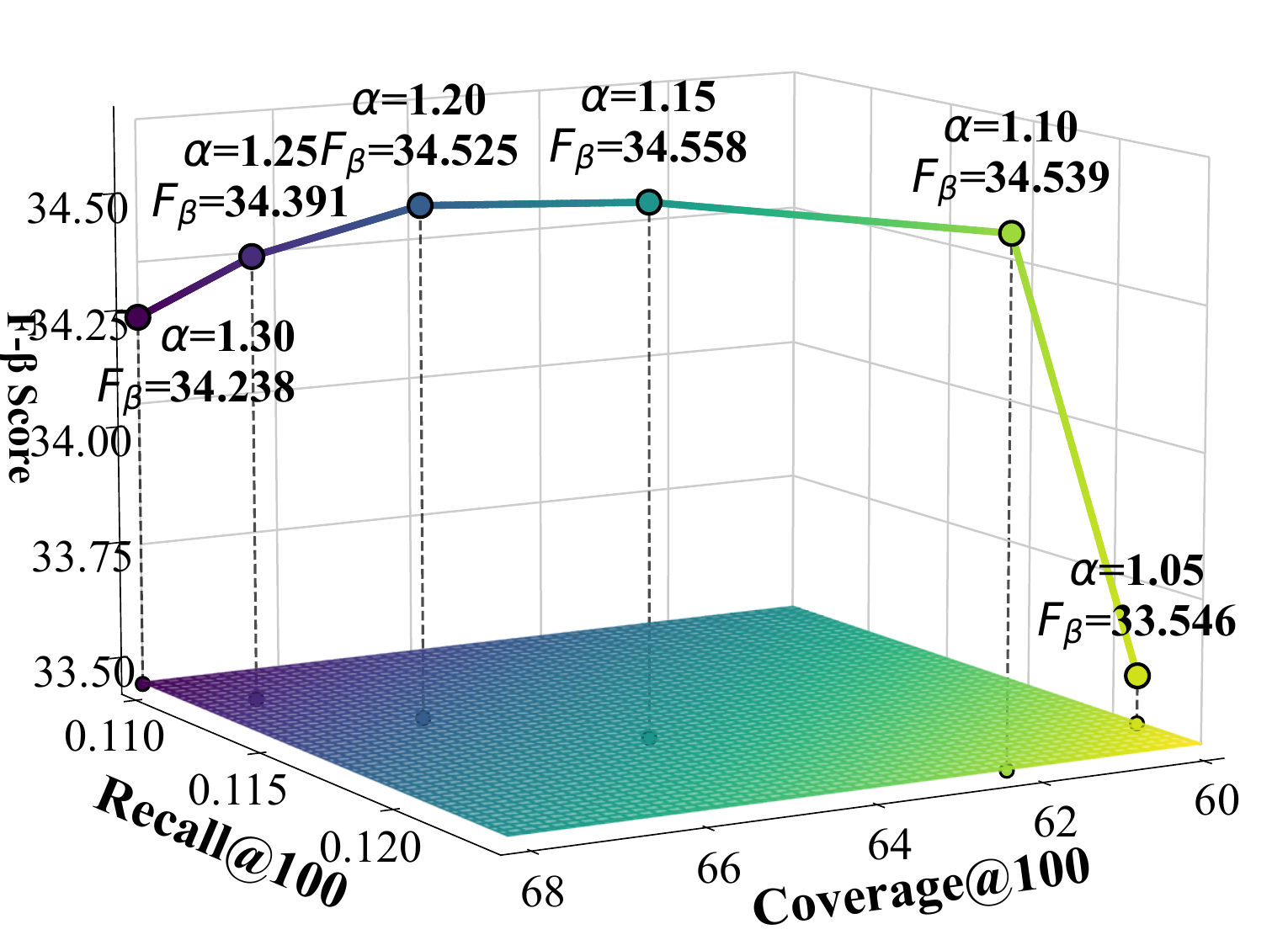}
    \caption{Toy-$\alpha$}
    \label{fig:toy_alpha}
\end{subfigure}%
\hfill%
\begin{subfigure}[b]{0.5\columnwidth}
    \centering
    \includegraphics[width=\textwidth]{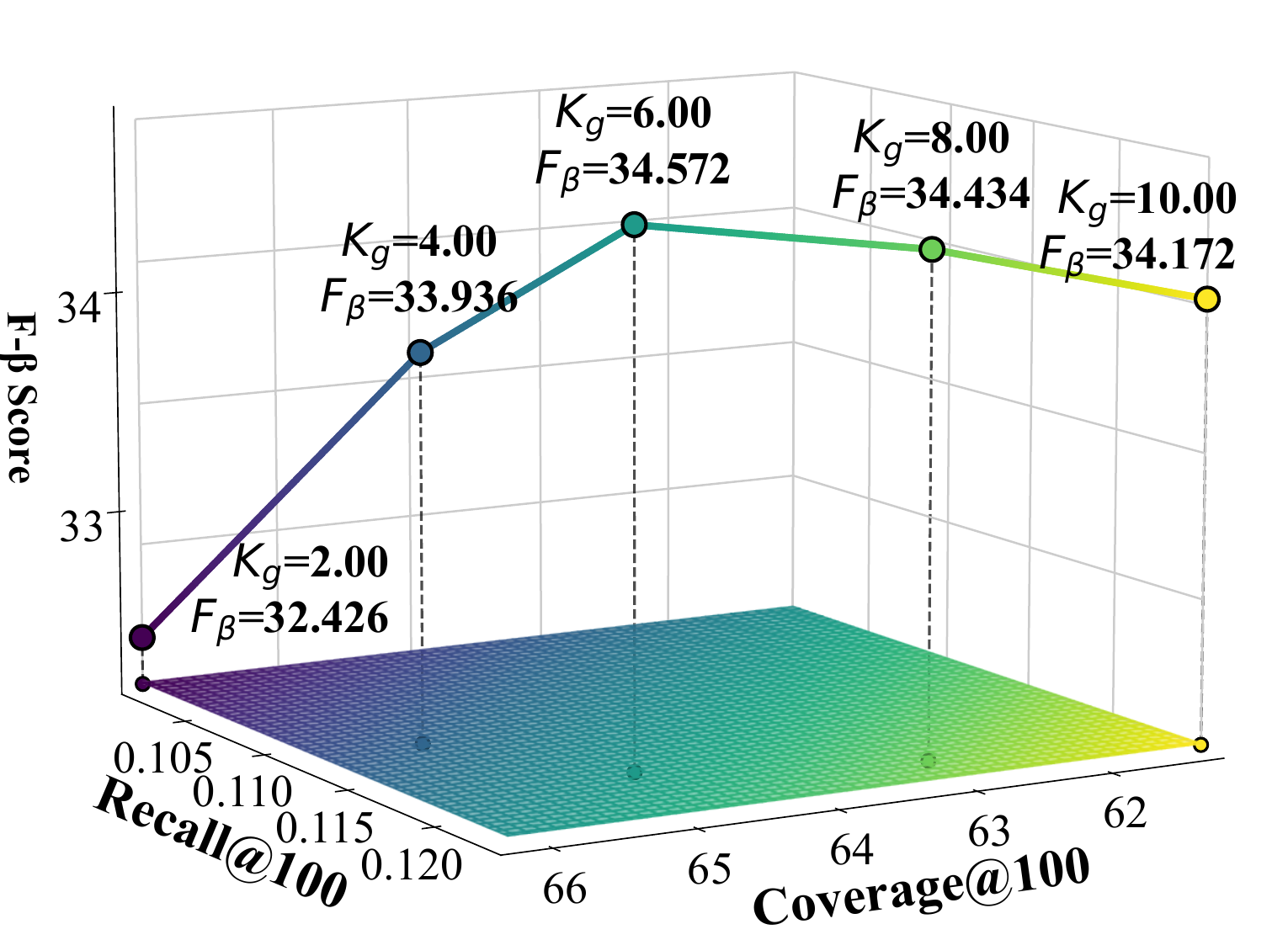}
    \caption{Toy-$K_g$}
    \label{fig:toy_kg}
\end{subfigure}

\caption{Parameter sensitivity on $\alpha$ and $K_g$}
\label{fig:parameter_sensitivity}
\end{figure}

In this subsection, we conduct experiments to study the impact of the scaling factor $\alpha$ and the global filtering coefficient $K_g$, as illustrated in Figure \ref{fig:parameter_sensitivity}. Under the recommendation list length setting of $\{100, 300\}$, the category-specific selection number $K_c$ achieves the best performance when set to 1 for all datasets. 
To simultaneously consider accuracy and diversity performance, we adopt the $F_\beta$-score as the evaluation metric, where the $\beta$ value can be adjusted according to specific requirements and application scenarios. The $F_\beta$-score is defined as:

\begin{equation}
F_\beta = (1 + \beta^2) \cdot \frac{\text{Coverage@100} \cdot \text{Recall@100}}{\beta^2 \cdot \text{Recall@100} + \text{Coverage@100}}
\end{equation}

Here, the $\beta$ values for the Beauty, TaoBao, and Toy datasets are set to 4, 20, and 15, respectively.


\paragraph{Effect of $\alpha$}$\alpha$ represents the intensity of counterfactual popularity enhancement. As $\alpha$ increases, potential interest items selected based on UACR are more likely to be successfully recommended, leading to a continuous rise in diversity metrics. This is because the two-stage filtering and aggregation strategy introduces more diversified candidate items. The accuracy metrics show a trend of first increasing and then decreasing, indicating that moderate $\alpha$ values can discover relevant potential interest items. However, when $\alpha$ becomes too large, it leads to recommending too many items that users are not interested in, thereby compromising recommendation quality. This phenomenon demonstrates the need to find an optimal balance between discovering potential diversified interests and maintaining recommendation relevance.

\paragraph{Effect of $K_g$}$K_g$ represents the number of candidate items in the global filtering process of the two-stage screening. As $K_g$ increases, the accuracy of the Beauty dataset shows a trend of first increasing and then decreasing, indicating that in small-scale dense datasets, moderately increasing the number of candidates helps discover relevant items, but too many candidates introduce noise and reduce accuracy. In contrast, the accuracy of TaoBao and Toy datasets continues to rise, showing that in large-scale sparse data environments, more candidate items can effectively alleviate data sparsity issues and significantly improve recommendation hit rates. However, the diversity metrics of all three datasets show a declining trend, because as the number of global candidates increases, mainstream items with high UACR scores occupy more recommendation positions, squeezing out the space for diverse candidates, thus sacrificing the diversity of recommendation results to some extent.
\subsection{Complexity Analysis. }

\begin{table*}[htbp]
\centering
\small
\resizebox{\textwidth}{!}{%
\begin{tabular}{l|*{6}{c}|cc}
\toprule
Dataset & EDUA & DGRec & DGCN & CPGRec & KG-Diverse & DivGCL & Ours & LightGCN \\
\midrule
Beauty & 4958.0593 & 201.0455 & 127.8954 & 130.6326 & 2018.9638 & 2331.0843 & \underline{97.5654} & \textbf{87.1020} \\
\midrule
TaoBao & 83792.6725 & 6018.0012 & 3443.7021 & 5168.6267 & 49538.6218 & 54355.1921 & \underline{2308.6280} & \textbf{2217.9973} \\
\midrule
Toy &117801.3664 & 8617.4631 &5594.5200 &7989.3924 & 8430.4206&53799.6355& \underline{2756.3432} & \textbf{2357.4224} \\
\bottomrule
\end{tabular}%
}
\caption{Total Training Time Comparison Across Different Datasets and Methods}
\label{tab:total_time_comparison}
\end{table*}

Our framework is built on the lightweight LightGCN backbone, with the insertion of two modules: UACR-Guided Causal Inference and Graph Refinement (CIGR) and UACR-Guided Candidate Selection and Counterfactual Exposure (CSCE). In LightGCN, user-item interactions are modeled through $L$-layer graph convolutions. Let $E$ denote the number of user-item edges, $d$ the embedding dimension, and $b$ the batch size. The time complexity of one forward pass is $O(L \times E \times d)$.


In the CIGR module, we compute UACR values only for item pairs with co-purchase relations. Let $\gamma$ denote the number of such co-purchase edges; the computation complexity is $O(\gamma)$, as these edges are typically sparse. As this process is executed once offline, the overhead is negligible. Afterward, we apply geometric truncation to prune the co-purchase graph to $\beta$ edges and perform $L_{II}$-layer graph convolution over this sparse graph. The time complexity is $O(L_{II} \times \beta \times d)$. The CSCE module is executed only once after training for offline inference, with a complexity of $O(U)$, which is negligible. The overall training complexity is $O(L \times E \times d + L_{II} \times \beta \times d)$. Therefore, the overall training complexity remains at the same level as LightGCN, since $\beta \ll \gamma$ and $\beta \ll E$.

To validate the training efficiency, we record the total time to convergence (measured in seconds) for each model. To ensure a fair comparison, we adopt an early stopping strategy based on Recall@100, where training is terminated if no improvement is observed for 10 consecutive epochs. Experimental results show that our method consistently outperforms all baselines, while achieving training efficiency comparable to LightGCN, as shown in Table \ref{tab:total_time_comparison}, demonstrating both its effectiveness and practicality.

\subsection{Performance Stability under Random Initialization}
\begin{figure}
   \centerline{\includegraphics[width=1.2\linewidth]{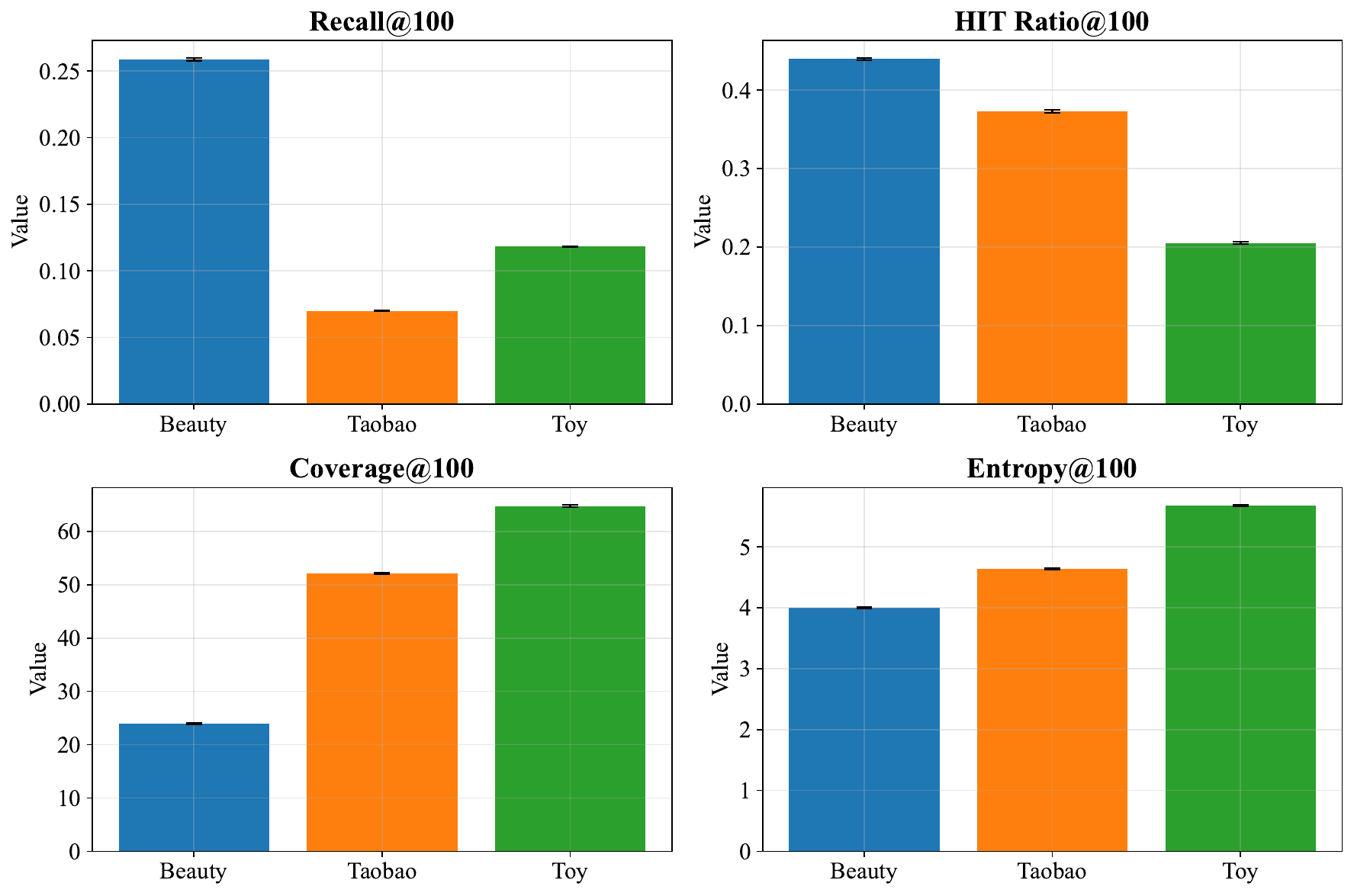}}
   \caption{Performance comparison with error bars across three datasets. The error bars represent the standard deviation over five independent runs with different random seeds.}
   \label{fig:performance_stability}
\end{figure}
To validate the robustness of Cadence, we conduct five independent runs for each experiment using different random seeds [2021, 2022, 2023, 2024, 2025], reporting the mean and standard deviation for each evaluation metric. As illustrated in Figure~\ref{fig:performance_stability}, Cadence consistently demonstrates superior performance with low variance across multiple independent executions, confirming its stability under various random initialization conditions.

To further establish the statistical significance of the performance improvements, we employ the Wilcoxon signed-rank test to compare Cadence against the top-performing baseline methods on each dataset: CPGRec on the Beauty dataset, DGRec on the TaoBao dataset, and DivGCL on the Toy dataset. The statistical analysis focuses on two key metrics: Recall@100 and Coverage@100. The test yields a statistic of 15.0 with a corresponding p-value of 0.0312, confirming that Cadence's improvements over the existing state-of-the-art methods are statistically significant at the 5\% significance level.

\section{Related Work}
\subsection{Counterfactual Recommendation}

In recent years, counterfactual recommendation has emerged as a key direction in causal recommender systems, aiming to correct selection and exposure biases introduced by historical logging policies. Early approaches mainly relied on propensity score modeling, such as inverse propensity scoring (IPS) \cite{agarwal2019general}, to achieve unbiased estimation from logged user feedback. With the rise of deep learning, recent work has explored integrating counterfactual reasoning with neural models. These approaches include generating counterfactual sequences in sequential recommendation \cite{wang2021counterfactual,zhang2021causerec}, synthesizing alternative interactions in knowledge graphs \cite{mu2022alleviating}, and modeling confounders through structural causal models \cite{chen2025causal}. These methods not only improve the accuracy and robustness of recommendation under biased data, but also provide theoretical foundations and practical solutions across various tasks, including causal disentanglement of multi-feedback signals in multi-behavior recommendation \cite{wang2023causal}, correction of contextual interference in session-based scenarios \cite{tang2025coder}, counterfactual explanation generation for interpretable recommendation \cite{li2024attention}, and mitigation of unfair exposure caused by user attributes or system biases in fair recommendation \cite{shao2024average}. 

\textbf{Discussion.} UCRS~\cite{wang2022user} uses user-issued controls and counterfactual inference to adjust recommendations by simulating changes in user features, aiming at controllable recommendation rather than explicit diversity optimization or causal item modeling. In contrast, our method enhances diversity from a causal perspective by constructing a deconfounded causal item graph and applying counterfactual exposure interventions to uncover under-exposed yet relevant items.
\subsection{Diversified Recommendation}
Diversity-aware recommendation aims to alleviate the issue of homogeneity in recommendation results and enhance overall user satisfaction \cite{ziegler2005improving}. Early approaches primarily relied on re-ranking strategies, such as Maximum Marginal Relevance (MMR) \cite{abdool2020managing,carbonell1998use,lin2022feature,peska2022towards} and Determinantal Point Process (DPP) \cite{chen2018fast,gan2020enhancing,huang2021sliding}, which optimize the recommendation list to promote item dissimilarity. With the advancement of graph neural networks \cite{tian2022reciperec}, more recent methods leverage high-order graph structures to capture complex user–item relationships for improved diversity, including category-aware sampling \cite{zheng2021dgcn}, submodular neighbor selection \cite{yang2023dgrec}, clustering-based candidate generation \cite{liu2024drgame}, and dynamic graph updates \cite{ye2021dynamic}. Moreover, the emergence of large language models (LLMs) has introduced new opportunities for diversity modeling, where controllable generation frameworks \cite{chen2025dlcrec} are developed to regulate category coverage and further enrich the diversity of recommendations. 

\textbf{Discussion.} Most existing approaches lack principled theoretical grounding and a causal perspective, limiting their capacity to generate genuinely diverse recommendations. Traditional co-occurrence-based methods are prone to confounding from item popularity and user attributes, leading systems to favor popular but homogeneous items while neglecting underexposed yet causally relevant ones. In our work, we employ causal inference to remove these spurious associations, recover true item dependencies, uncover latent user interests, and enhance diversity without compromising relevance.

\section{Detailed Proof of Theorem: Norm-Popularity Correlation}

This section provides the complete mathematical proof of the norm–popularity correlation theorem.

\subsection*{A.1 Notation and Sampling Assumptions}

\begin{itemize}
\item $\mathcal U,\mathcal I$: user set and item set;
\item $A$: the random-walk normalised adjacency matrix with $\|A\|_2 \le 1$;
\item $M = \sum_{k=0}^{K} \alpha_k A^k$: LightGCN propagation operator with $\alpha_k \ge 0$ and $\sum_k \alpha_k = 1$;
\item $n_i$: number of positive interactions for item~$i$ (``popularity''), and $|D| = \sum_i n_i$;
\item $\eta,\lambda$: SGD step size $\eta$ and $\ell_2$ regularisation coefficient $\lambda$;
\item $\mathbf e_x^{(\ell)}$: embedding of node $x$ at layer $\ell$; the final embedding is defined as $\displaystyle \mathbf e_x = \sum_{k=0}^{K} \alpha_k \,\mathbf e_x^{(k)}$;
\item $\Delta = \langle \mathbf e_u,\mathbf e_i \rangle - \langle \mathbf e_u,\mathbf e_j \rangle$;
\item $\sigma(\cdot)$: sigmoid function.
\end{itemize}

\textbf{Default sampling strategy.}  At each training step we randomly draw one positive interaction $(u,i)$ and then uniformly sample a negative item $j$ from the set of items not interacted by~$u$; hence
\begin{equation}
\Pr[i] = \frac{n_i}{|D|}.
\label{eqA:prob}
\end{equation}

\subsection*{A.2 Full Chain-Rule Derivation of $\partial\ell/\partial\mathbf e_i^{(0)}$}

The BPR loss is
\begin{equation}
\ell = -\log \sigma(\Delta), \qquad
\frac{\partial \ell}{\partial \Delta} = 1-\sigma(\Delta).
\end{equation}

\paragraph{Gradient with respect to the final embeddings.}
\begin{equation}
\frac{\partial \ell}{\partial \mathbf e_i} = (1-\sigma(\Delta))\,\mathbf e_u,
\end{equation}
\begin{equation}
\frac{\partial \ell}{\partial \mathbf e_j} = -(1-\sigma(\Delta))\,\mathbf e_u,
\end{equation}
\begin{equation}
\frac{\partial \ell}{\partial \mathbf e_u} = (1-\sigma(\Delta))(\mathbf e_j-\mathbf e_i).
\end{equation}

\paragraph{Backpropagation to $\mathbf e_i^{(0)}$}  
Since $\mathbf e = M\mathbf e^{(0)}$ (LightGCN only performs linear propagation), we have
\begin{equation}
\frac{\partial\mathbf e_i}{\partial\mathbf e_i^{(0)}} = M_{ii},
\end{equation}
\begin{equation}
\frac{\partial\mathbf e_u}{\partial\mathbf e_i^{(0)}} = M_{ui},
\end{equation}
\begin{equation}
\frac{\partial\mathbf e_j}{\partial\mathbf e_i^{(0)}} = 0.
\end{equation}
Therefore
\begin{equation}
\frac{\partial\ell}{\partial\mathbf e_i^{(0)}}
=(1-\sigma(\Delta))
\bigl[M_{ii}\mathbf e_u+M_{ui}(\mathbf e_j-\mathbf e_i)\bigr].
\label{eqA:grad_full}
\end{equation}

\paragraph{Component along $\hat{\mathbf e}_i^{(0)}$ direction}
Let $\theta_{ui}$ be the angle between $\mathbf e_u$ and $\mathbf e_i^{(0)}$,
$\rho_i:=\|\mathbf e_i\|_2/\|\mathbf e_u\|_2$, then
\begin{equation}
g_{ui}^{\parallel}
=(1-\sigma(\Delta))
\bigl[M_{ii}\cos\theta_{ui}+M_{ui}\cos\theta_{ui}-M_{ui}\rho_i\bigr]
\|\mathbf e_u\|_2.
\label{eqA:g_parallel}
\end{equation}

\subsection*{A.3 Single-step Norm Increment}

SGD update with small learning rate for stability, $T\eta\!\ll\!1$, ignoring $O(\eta^2)$ terms
\begin{equation}
\mathbf e_i^{(0)}\leftarrow
\mathbf e_i^{(0)}-\eta[\mathbf g+2\lambda\mathbf e_i^{(0)}],
\quad
\mathbf g=\frac{\partial\ell}{\partial\mathbf e_i^{(0)}}.
\end{equation}
Let $g_\parallel=\hat{\mathbf e}_i^{(0)\top}\mathbf g$, then
\begin{equation}
\Delta\|\mathbf e_i^{(0)}\|_2
=-\eta g_\parallel-2\eta\lambda\|\mathbf e_i^{(0)}\|_2+O(\eta^2).
\label{eqA:delta_norm}
\end{equation}

Substituting \eqref{eqA:g_parallel} and taking the conditional expectation over $(u,i,j)$,
in large-scale datasets $\bar\kappa$ and $\beta$ vary minimally across items, so we approximate them as constants
\begin{equation}
\bar\kappa := \mathbb{E}_{u|i}[(1-\sigma(\Delta))(M_{ii}\cos\theta_{ui}+M_{ui}\cos\theta_{ui})],
\end{equation}
\begin{equation}
\beta := \mathbb{E}_{u|i}[(1-\sigma(\Delta))M_{ui}/\|\mathbf e_u\|_2].
\end{equation}
Since user embedding norms have no significant correlation with individual item popularity, in the statistical sense of large datasets we can treat $\mathbb{E}_{u|i}[\|\mathbf e_u\|_2]$ as a constant
\begin{equation}
\mathbb{E}[\Delta\|\mathbf e_i^{(0)}\|_2]
=\eta\!\Bigl[(\bar\kappa-\beta\|\mathbf e_i\|_2)\frac{n_i}{|D|}
          -2\lambda\|\mathbf e_i^{(0)}\|_2\Bigr].
\label{eqA:exp_delta}
\end{equation}

\subsection*{A.4 Fixed-Point Analysis and Linear Relationship}

Define $\mu_i := \mathbb{E} \|\mathbf e_i\|_2$. At steady state $\mathbb{E}[\Delta \|\mathbf e_i^{(0)}\|_2] = 0$, i.e.
\begin{equation}
(\bar\kappa - \beta \mu_i) \frac{n_i}{|D|} = 2\lambda \mu_i.
\end{equation}
Solving yields
\begin{equation}
\boxed{\mu_i = \frac{\bar\kappa}{2\lambda + \beta}\,\frac{n_i}{|D|}}\quad (\beta \ge 0).
\label{eqA:fixedpoint}
\end{equation}
Because the dependence of $\bar\kappa$ and $\beta$ on $i$ is extremely weak, $\mu_i$ grows linearly with $n_i$ and the proportionality constant is essentially item-independent.

\subsection*{A.5 High-Probability Norm Upper Bound}

\paragraph{Hard Bound. }
Let $\eta \le 1/(4\lambda)$ and set $C = \bar\kappa/(2\lambda)$ and $c = 2\eta(\bar\kappa + \lambda C)$. Then for any $\epsilon>0$ and $T \in \mathbb N$,
\begin{equation}
\Pr\!\Bigl( \sup_{t \le T} \|\mathbf e_{i,t}^{(0)}\|_2 > C + \epsilon \Bigr)
\le \exp\!\bigl(-\epsilon^{2}/(4Tc^{2})\bigr).
\end{equation}

\paragraph{Proof.}
Let $Z_t = \|\mathbf e_{i,t}^{(0)}\|_2 - C$. From~\eqref{eqA:delta_norm} and the bound $|g_{\parallel}| \le \bar\kappa \|\mathbf e_u\|_2 + |M_{ui}| \|\mathbf e_j\|_2 \le \bar\kappa C$, we have $|Z_{t+1} - Z_t| \le c$. Moreover, $\mathbb{E}[Z_{t+1}-Z_t \mid \mathcal F_t] \le 0$, so $\{Z_t\}$ is a supermartingale. The maximal-difference form of the Azuma–Hoeffding inequality then gives the claimed bound. 

The parameters $\bar\kappa$ and $C$ vary little across items, so the bounded-difference condition holds uniformly for all~$i$.

\subsection*{A.6 Theorem Statement (Uniform Negative Sampling)}

Combining~\eqref{eqA:prob}, \eqref{eqA:fixedpoint} and the above bound, we obtain
\begin{equation}
\|\mathbf e_i\|_2 = \Theta(n_i)
\quad\text{and}\quad
\Pr\bigl( \|\mathbf e_i\|_2 > \|\mathbf e_j\|_2 \bigr) \ge 1 - e^{-\Omega(1)}.
\end{equation}

\noindent This completes the proof.

\bibliography{aaai2026}

\end{document}